\documentstyle[referee]{laa}

\begin{document}
\thesaurus{07  %A&A Section: Solar system
	   (07.09.1; % Interplanetary medium
	    07.13.1;  % Meteoroids
	   )}

\title{Real Dust Particles and Unimportance of the Poynting-Robertson Effect}
\author{M.~Kocifaj $^{1)}$ and J.~Kla\v{c}ka $^{2)}$}
\institute{Institute of Astronomy,
   Slovak Academy of Sciences, D\'{u}bravsk\'{a} cesta 9, \\
   842~28 Bratislava, Slovak Republic $^{1)}$ \\
Institute of Astronomy,
   Faculty for Mathematics and Physics, Comenius University \\
   Mlynsk\'{a} dolina, 842~15 Bratislava, Slovak Republic $^{2)}$}
\date{}
\maketitle

\begin{abstract}
The importance of the Poynting-Robertson effect on the motion of
interplanetary dust particles is discussed. Precise numerical calculations
for real dust particle show that condition for the validity of the
Poynting-Robertson effect is not fulfilled. The interaction of the
(solar) electromagnetic radiation with really shaped dust particle is different
from that which yields the Poynting-Robertson effect. The magnitude of the
Poynting-Robertson effect's deceleration term is in one to two orders in
magnitude (it depends on particle's size) less important than terms
corresponding to nonforward (or, nonbackward) scattering.

%\keywords{celestial mechanics, stellar dynamics, dust, radiation scattering}

\end{abstract}

\section{Introduction}
The Poynting-Robertson effect (P-R effect) (Robertson 1937; Kla\v{c}ka 1992a --
the most complete form of the P-R effect) is generally considered to be the
real effect which causes inspiralling of interplanetary dust particles (IDPs),
meteoroids, toward the Sun. (Other, more simple correct derivations may be found
in Kla\v{c}ka's papers: 1992b, 1993a, 1993b). The most general case of the validity
of the P-R effect requires that Eq. (120) (or, Eq. (122) for the moving
particle) in (Kla\v{c}ka 1992a) holds. This may not be the case for the real
nonspherical particle, as it was discussed in Kla\v{c}ka (1993c, 1993d (some
numerical errors, which may be easily found, are in the last section; moreover,
real particle should rotate around one axis -- axis of rotation; the important
result is that the significance of the transversal components increases with
increasing porousity and reflectivity of dust particle), 1994b) and
applied in Kla\v{c}ka (1994a).

General equation of motion of IDP in terms of optical properties was presented
by Kla\v{c}ka and Kocifaj (1994) -- the paper does not present any quantitative
calculation for real particles.

The aim of this paper is to present quantitative results for real stationary
(not moving around the Sun) IDP.

\section{More Detailed Qualitative Discussion}
Interplanetary dust particles are different in size, chemical composition, shape and
physico-optical properties. The light scattering by such particles predetermines changes
of their motion in the space. This fact is expressed by well-known radiation pressure,
which was notoriously based on spherical target assumption. Scattering diagram for
spheres shows the radial symmetry in Sun-particle coordinated system. The only
forward and backward scattering efficiences are important in this case. However, any
irregularity of the particle shape will produce certain momentum in perpendicular
projections to the direction of light propagation. This may be caused also by inhomogeneity
of particle chemical composition (or particle density). Particle shape specificity
(unconcavities, cavities,...) play dominant role in formation of light scattering
diagram. Several studies on nonspherical particles start with model types of particles
such as ellipsoids, cylinders, rectangular targets. However, this approach must allways
take into account a certain degree of particle shape symmetry, which is responsible for
decreasing of light intensity scattered to the perpendicular directions. Really shaped
particle is therefore the best candidate to make an appropriate conclusions. We are
using the computer model of real cosmic dust particle U2015B10
(Clanton {\it et. al}, 1984).

\section{Model particle\label{particle}}
The computer model of the really shaped particle is practically identical with the
catalogued cosmic dust particle U2015B10. The only total size (radius of an equivalent
sphere) was modified to reach required conditions for theoretical study (to enable
perform the calculation of scattering effects by wide particle population). Computer
model was based on scanning of particle by electron microscope. The particle consists
mainly from Mg, Al, S, Ca and Fe. We have also found that enstatite, constitution of
which is near to $Mg_{0.8}Fe^{II}_{0.2}SiO_{3}$, dominates in particle composition.
From the morphological point of view, it is remarkable that the U2015 B10 sample
has probably a cavity in its center (for the further details see Kapi\v{s}insk\'{y}
{\it et al.} , 1995). Since the mineral enstatite is dominant in composition of U2015B10,
the mean refractive index of the particle can be, in sufficient accuracy, fitted
by the enstatite optical properties ($n_{r} = 1.735 - 0.26 \lambda$, $n_{i} < 10^{-4}$).
The values $n_{r}$ and $n_{i}$ represent the real and imaginary part of particle
refractive index, respectively. Detailed description of the model and its setup
characteristics for calculation are discussed in (Kocifaj {\it et. al}, 1999).

\section{Calculation method\label{calculation}}
The characteristics of radiation scattered by examined cosmic dust particle were calculated
using the so-called Discrete Dipole Approximation (Draine, 1988). Particle effective
radius was $0.5~\mu m$ while the wavelength of an incident radiation cover the visible
spectrum and near infrared spectral band (i.e., the spectral range which corresponds
to the maximum energy distribution in solar spectrum). The mean direction of
propagation of the scattered radiation is characterized by vector $\vec{g}$
\begin{equation}
\vec{g} = ~ < ~\cos \theta_{s} ~> ~ \vec{x} ~+~
	    < ~\sin \theta_{s} ~\cos \phi_{s} ~> ~\vec{y} ~+~
	    < ~\sin \theta_{s} ~\sin \phi_{s} ~> ~\vec{z} ~,
\end{equation}
where $\theta_{s}$ and $\phi_{s}$ represent the scattering angle and azimuth measured
around the axis $\vec{x}$ characterizing the direction of propagation of the
incident radiation. The radiative energy scattered to the directions perpendicular to
$\vec{x}$ (i.e., $\vec{y}$ and $\vec{z}$) are expressed by the last two terms. They are
corresponding to the parameters $G_{1}$ and $G_{2}$, while $G_{0}$ belongs to the first
term, using the formalism utilized in Kocifaj and Kapi\v{s}insk\'{y} (1997). The P-R effect
may be unappropriate if ratios $G_{x}/G_{0}$ ( $x =$ 1, 2 ) are greater than $\approx
10^{-4}$. Rotation of the particle in the space can decrease the ratios by averaging
over the whole solid angle. Average value of the parameter $G$ can be expressed as follows:
\begin{equation}\label{2}
< ~G ~> ~= \frac{1}{8~\pi^{2}} ~\int_{0}^{2 ~\pi} ~d \beta \int_{-1}^{1} ~
	   d ~\cos \theta ~ \int_{0}^{2~ \pi} ~
	   G ( \beta, \theta, \phi) ~d \phi,
\end{equation}
where $\beta$, $\theta$ and $\phi$ are angles which represent the particle orientation
in the lab-frame (Draine and Flatau, 1996). The target is oriented such that the polar
angles $\theta$ and $\phi$ specify the direction of a selected unit vector in the particle
relative to the incident radiation. The target is assumed to be rotated around this unit
vector by an angle $\beta$.

Averaging in wavelengths is defined by the following equation:
\begin{equation}\label{3}
< ~\vec{g}_{integr} ~> ~= \int_{\lambda_{1}}^{\lambda_{2}} ~I_{0} ( \lambda )
			  ~ \vec{g} ~ d \lambda ~,
\end{equation}
where $I_{0} ( \lambda )$ is intensity of the incident radiation.

\section{Results\label{results}}
The mean values of individual components of the vector $\vec{g}$ depend on the
particle orientation in the laboratory frame. Averaged values of ratios $G_{x}/G_{0}$
for rotating particle are functions of orientation of the rotation plane
(or axis of rotation). The examined ratios will be very small for particles
with a relatively small asymmetry when cross section varies slowly during rotation
(e.g., axis of rotation is parallel to the direction of the incident radiation).
We have studied rotation in three planes, when two angles from $\beta$, $\theta$,
or $\phi$ are constant and one of them varies over whole possible range. It is
evident that rotation with the variable angle $\phi$ brings very small values
of ratios $G_{x}/G_{0}$. This is caused by practically constant cross section
in all particle positions (the changes of angle $\phi$ express the precession
of the particle in our case). Results of calculation for monochromatic radiation
and polychromatic radiation are presented in Tables 1 and 2. Rotation of the particle
derived from the monotonous change of the angle $\beta$, or $\theta$, brings the
fluctuation of the particle cross section. This fact results in an asymmetry of the
radiation scattering diagrams. Such a case indicates an increase of ratios $G_{x}/G_{0}$.
Calculated data summarized in Table 1 confirm high value of these ratios (about 0.02-0.07).

\begin{table}[t1]
\begin{center}
\label{t1}
\caption{Spectral radiation characteristics calculculated at selected wavelengths}
\begin{tabular}{c} \\
\hline
\hline
rotation of the particle derived from monotonous change of angle $\beta$ \\
\begin{tabular}{rrrrrr}
\hline
$\lambda~[\mu~m]$ & $G_{0}$ & $G_{1}$ & $G_{2}$ & $G_{1}/G_{0}$ & $G_{2}/G_{0}$ \\
\hline
0.50 & 1.231 & 0.094 & -0.003 & 0.076 & -0.003 \\
0.56 & 1.219 & 0.097 & 0.003 & 0.080 & 0.002 \\
0.64 & 1.205 & 0.094 & 0.001 & 0.078 & 0.001 \\
0.75 & 1.156 & 0.098 & -0.003 & 0.085 & -0.003 \\
0.90 & 1.056 & 0.082 & -0.002 & 0.078 & -0.002 \\
\hline
\hline
\end{tabular}
\end{tabular}
\begin{tabular}{c} \\
\hline
\hline
rotation of the particle derived from monotonous change of angle $\theta$ \\
\begin{tabular}{rrrrrr}
\hline
$\lambda~[\mu~m]$ & $G_{0}$ & $G_{1}$ & $G_{2}$ & $G_{1}/G_{0}$ & $G_{2}/G_{0}$ \\
0.50 & 1.270 & 0.030 & -0.042 & 0.023 & -0.033 \\
0.56 & 1.301 & 0.036 & -0.054 & 0.028 & -0.042 \\
0.64 & 1.288 & 0.049 & -0.015 & 0.038 & -0.011 \\
0.75 & 1.279 & 0.047 & 0.046 & 0.037 & 0.036 \\
0.90 & 1.128 & 0.029 & 0.087 & 0.026 & 0.077 \\
\hline
\hline
\end{tabular}
\end{tabular}
\begin{tabular}{c} \\
\hline
\hline
rotation of the particle derived from monotonous change of angle $\phi$ \\
\begin{tabular}{rrrrrr}
\hline
$\lambda~[\mu~m]$ & $G_{0}$ & $G_{1}$ & $G_{2}$ & $G_{1}/G_{0}$ & $G_{2}/G_{0}$ \\
0.50 & 1.477 & -0.000 & 0.000 & 0.000 & 0.000 \\
0.56 & 1.454 & 0.000 & 0.000 & 0.000 & 0.000 \\
0.64 & 1.366 & 0.000 & -0.000 & 0.000 & 0.000 \\
0.75 & 1.315 & 0.000 & 0.000 & 0.000 & 0.000 \\
0.90 & 1.123 & -0.000 & -0.000 & 0.000 & 0.000 \\
\hline
\hline
\end{tabular}
\end{tabular}
\end{center}
\end{table}

\begin{table}[t2]
\begin{center}
\label{t1}
\caption{Integral values of radiation characteristics $G$ averaged over visible spectrum}
\begin{tabular}{lrrrrr} \\
\hline
\hline
$angle$ & $G_{0}$ & $G_{1}$ & $G_{2}$ & $G_{1}/G_{0}$ & $G_{2}/G_{0}$ \\
\hline
$\beta$ & 0.670 & 0.054 & -0.000 & 0.080 & -0.001 \\
$\theta$ & 0.720 & 0.023 & 0.001 & 0.033 & 0.001 \\
$\phi$ & 0.770 & -0.000 & 0.000 & 0.000 & 0.000 \\
\hline
\end{tabular}
\end{center}
\end{table}

Integral values presented in Table 2 were obtained by integrating $\vec{g}$ over the
studied spectral band from $\lambda$~=~0.5~$\mu m$ to $\lambda$~=~0.9~$\mu m$. Radius
of the particle was 0.5 $\mu m$. The ratio $G_{x}/G_{0}$ is at level $10^{-2}$ which
is in two magnitudes greater than the limit value of $10^{-4}$. The results for the 5.0
micron particle (radius~=~5.0 $\mu m$) are approximately 20-50 times less than values
given above, i.e., the ratio is still not less than $10^{-4}$.
It means, the submicron and small micron particles violates the limit value of $10^{-4}$.
Scattering diagrams presented in Figures 1-3 show the distribution of the energy
scattered in the halfsphere outward the Sun (i.e. in direction of incident radiation).
Wavelength of the incident radiation was 1.0 $\mu m$ in this case. However, the isolines
are asymmetric for $\theta$ and $\beta$ also after integration over all particle
orientations during rotation and over all used wavelengths. It is important result.
It characterizes the real influence of particle shape on the resulting motion
of the particle in the space. The values of individual isolines vary only over two
magnitudes -- this is the reason why the ratios $G_{x}/G_{0}$ are so important.
Figure 3 presents the resulting scattering diagram after integration over all particle
positions derived from changes of angle $\phi$. This diagram is completely identical to
the scattering diagram for the sphere. It documents that the rotation of the particle may
significantly decrease the relatively high values of the examined ratios (Table 1).
It also documents the level of numerical precision of the calculation. Results for the
other types of rotation (expressed in terms of angles $\beta$ and $\theta$) show that
one component of $G_{x}/G_{0}$ is preferred. The ratio $G_{1}/G_{0}$ (axis $90^{o}-270^{o}$)
is greater than $G_{2}/G_{0}$ (axis $0^{o}-180^{o}$) for the rotation derived from changes
of the angle $\beta$ (Fig. 2) -- partial symmetry of the scattering diagram over the
axis ($0^{o}-180^{o}$) is evident. The importance of one of the components $G_{x}/G_{0}$
predetermines the acceleration of the particle in the defined direction. Moving the particle
to the other position relatively to the Sun can bring a change of the preferred direction
of motion due to a possible new lab-frame conditions. The particle may be accelerated only
during a limited time. Another position of the particle with respect to the Sun may
decelerate its motion.

\section{Conclusion}
We have shown that real dust particles may exhibit important nonforward and
nonbackward scattering efficiences in comparison with the result when the P-R
effect holds. Therefore, the P-R effect cannot be simply applied to the study
of a particle motion in space without precise physico-optical analysis.

\acknowledgements
The paper was supported by the Scientific Grant Agency VEGA (grants Nos. 1/4304/97,
1/4303/97, and 1/4174/97). The authors are also very grateful to Bruce Draine and
Piotr Flatau for their DDSCAT code.

\section*{Figure caption}
\begin{itemize}
\item [Fig. 1.] Scattering diagram for the particle rotation derived from monotonous change
of the angle $\theta$. Source of radiation (Sun) is behind the centre of the picture.
The azimuth angle $\phi_{s}$ is displayed on the border of the diagram.
\item [Fig. 2.] Scattering diagram for the particle rotation derived from monotonous change
of the angle $\beta$. Source of radiation (Sun) is behind the centre of the picture.
The azimuth angle $\phi_{s}$ is displayed on the border of the diagram.
\item [Fig. 3.] Scattering diagram for the particle rotation derived from monotonous change
of the angle $\phi$. Source of radiation (Sun) is behind the centre of the picture.
The azimuth angle $\phi_{s}$ is displayed on the border of the diagram.
\end{itemize}
\end{document}